\documentclass{aa}
\usepackage{graphicx}
\usepackage{txfonts}
\begin{document}

\title{The influence of chemical composition on the properties of Cepheid
  stars\\
  I - Period-Luminosity relation vs iron abundance\thanks{Based on observations
  made with ESO Telescopes at La Silla and Paranal Observatories under proposal
  ID 66.D-0571}}
\author{M. Romaniello\inst{1}
  \and F. Primas\inst{1}
  \and M. Mottini\inst{1}
  \and M. Groenewegen\inst{2}
  \and G. Bono\inst{3}
  \and P. Fran\c{c}ois\inst{4}
}
\offprints{M. Romaniello,\\ \email{mromanie@eso.org}}

\institute{European Southern Observatory, Karl-Schwarzschild-Strasse 2,
    D--85748 Garching bei M\"unchen, Germany
  \and
    Instituut voor Sterrenkunde, PACS-ICC, Celestijnenlaan 200B,
      B--3001 Leuven, Belgium
  \and
    INAF-Osservatorio Astronomico di Roma, via di Frascati 33, I--00040
      Monte Porzio Catone, Italy
  \and
    Observatoire de Paris-Meudon, GEPI, 61 avenue de l'Observatoire,
      F--75014 Paris, France
}

\date{Received October 8, 2004; accepted November 20, 2004}

\abstract{
We have assessed the influence of the stellar iron content on the
Cepheid Period-Luminosity ($PL$) relation by relating the $V$ band
residuals from the Freedman et al~(\cite{fre01}) $PL$ relation to
[Fe/H] for 37 Galactic and Magellanic Clouds Cepheids. The iron
abundances were measured from FEROS and UVES high-resolution and
high-signal to noise optical spectra. Our data indicate that the stars
become fainter as metallicity increases, until a plateau or turnover
point is reached at about solar metallicity. Our data are incompatible
with both no dependence of the $PL$ relation on iron abundance, and
with the linearly decreasing behavior often found in the literature
(e.g.  Kennicutt et al \cite{ken98}, Sakai et al \cite{sak04}). On the
other hand, non-linear theoretical models of Fiorentino et
al~(\cite{fio02}) provide a fairly good description of the data.
 }

\titlerunning{The Cepheid PL relation vs [Fe/H]}
\authorrunning{M. Romaniello et al}
\maketitle

\section{Introduction\label{sec:intro}}
Ever since the work of Edwin Hubble, the Cepheid Period-Luminosity
($PL$) relation is a fundamental tool in determining Galactic and
extragalactic distances. In spite of its paramount importance, to this
day we still lack firm theoretical and empirical assessment on whether
or not chemical composition has any significant influence on it.

Theoretical pulsational models by different groups lead to markedly
different results. On the one side computations based on \emph{linear
models} (e.g. Chiosi et al \cite{chi92}, Sandage et al \cite{san99},
Baraffe \& Alibert \cite{bar01}) suggest a mild dependence of the $PL$
relation on chemical composition. The predicted change at $\log(P)=1$
is less than 0.1 mag at all wavelengths between the $V$ and $K$ bands
for a change in metallicity from $Z=0.004$, representative of the
Small Magellanic Cloud (SMC), to 0.02, typical of Galactic Cepheids at
the Solar circle. This result is challenged by the outcome of the
\emph{non-linear convective models} (e.g. Bono et al \cite{bon99},
Caputo et al \cite{cap00}), which find that both the slope and the
zeropoint of the Period-Luminosity relation depend significantly on
the adopted chemical composition. Again for $\log(P)=1$ and the same
variation in metallicity as above, they predict a change as large as
0.4 magnitudes in $V$, 0.3 magnitudes in $I$ and $0.2$ magnitudes in
$K$. Moreover, the change is such that metal-rich Cepheids are {\it
fainter} than metal-poor ones, again at variance with the results of
Baraffe \& Alibert (\cite{bar01}). Recent calculations by Fiorentino
et al (\cite{fio02}), also based on non-linear models, indicate that
the $PL$ relation also depends on the helium abundance.

Observationally, indirect measurements in external galaxies from
secondary abundance indicators tend to find that metal-rich Cepheids
are {\it brighter} than metal-poor ones, albeit with disappointingly
large range of quoted values anywhere between 0 (Udalski et al
\cite{uda01}, Ciardullo et al \cite{cia02}) and
$-0.9~\mathrm{mag}/\mathrm{dex}$ (Gould \cite{gou94}). For lack of
better evidence, a marginally significant dependence of
$-0.2\pm0.2~\mathrm{mag}/\mathrm{dex}$ was adopted in the final paper
of the {\sc hst} Key Project on $\mathrm{H}_0$ (Freedman et al
\cite{fre01}). This was largely based on the results of Kennicutt et
al (\cite{ken98}) from oxygen abundances of H{\sc ii} regions in two
Cepheid fields in M101 ($-0.24\pm0.16~\mathrm{mag}/\mathrm{dex}$) and
in 10 Cepheid galaxies with Tip of the Red Giant Branch distances
($-0.12\pm0.08~\mathrm{mag}/\mathrm{dex}$). This latter method was
recently applied by Sakai et al (\cite{sak04}) to a larger sample of
17 galaxies yielding $-0.24\pm0.05~\mathrm{mag}/\mathrm{dex}$. Storm
et al (\cite{sto04}) assumed a metallicity difference between the
Milky Way and the Small Magellanic Cloud of $0.7~\mathrm{dex}$ to
derive a slope of $-0.21\pm0.19~\mathrm{mag}/\mathrm{dex}$ from 5 SMC
and 34 Galactic Cepheids for which they measured the distances with a
Baade-Wesselink technique.  Recently Groenewegen et al (\cite{gro04b})
have assembled from the literature and homogenized a large sample of
Cepheid stars with both distance and metallicity determinations to
derive a metallicity effect of
$-0.27\pm0.08~\mathrm{mag}/\mathrm{dex}$ in the zero point in VIWK
bands.

Compared to these previous studies, the novelty of our approach
consists in measuring directly the chemical composition of Cepheid
stars with know a distance, without relying on proxies such as oxygen
nebular abundances derived from spectra of H~II regions at the same
galactocentric distance of the Cepheid field (e.g. Kennicutt et al
\cite{ken98}), or secondary distance indicators like the Tip of the
Red Giant Branch (e.g. Sakai et al \cite{sak04}).

The paper is organized as follows. The observations are presented in
section~\ref{sec:obs}, together with a brief description of the [Fe/H]
measurements. In section~\ref{sec:plz} the dependence of the $PL$
relation on [Fe/H] is derived. Finally, section~\ref{sec:disc}
contains the discussion and the conclusions.

\section{Observations and data reduction\label{sec:obs}}
In order to achieve an extended coverage in metallicity, we have
observed Cepheid stars in three environments known to have
significantly different mean chemical compositions: the Solar
Neighborhood, the Large Magellanic Cloud (LMC) and the Small
Magellanic Cloud.

The spectra of the 13 Galactic stars were obtained with the FEROS
instrument (Pritchard \cite{pri04}\footnote{see also\\
\texttt{http://www.ls.eso.org/lasilla/sciops/2p2/E2p2M/FEROS}}) at the
ESO 1.5m telescope on Cerro La Silla. The spectral resolution is
48,000 and the signal-to-noise ratio is about 70 to 150, depending on
the brightness of the target and on the spectral range. The 12
Cepheids in the LMC and the 12 Cepheids in the SMC were observed with
the UVES spectrograph (Kaufer et al \cite{kau04}\footnote{see also
\texttt{http://www.eso.org/instruments/uves}}) at the VLT-Kueyen
telescope on Cerro Paranal. For them the resolution is about 40,000
and the signal-to-noise ratio about 50 to 70. The 2-D raw spectra were
run through the respective instrument pipelines, yielding to 1-D
extracted, wavelength calibrated and rectified spectra. The
normalization of the continuum was refined with the IRAF task
\emph{continuum}.  The 1-D spectra were corrected for heliocentric
velocity using the \emph{rvcorr} and \emph{dopcor} IRAF task. This
latter task was also used to apply the radial velocity correction,
which was derived from 20 FeI, FeII e MgI lines.

Selected characteristics of the programme stars are listed in Table~1.

\subsection{Metallicity determination\label{sec:detz}}
The complete description of the method we have used to derive the
metallicity for our programme stars will be presented in a forthcoming
paper (Mottini et al, in preparation). Here we just summarize the
salient points.

The iron content of the Cepheids was derived from the equivalent
widths (EWs) of 150-200 Fe~I and 10-15 Fe~II unblended lines. The EWs
themselves were measured semi-interactively with a software developed
by one of us (PF, \emph{fitline}). For about 15\% of the lines it was
necessary to use the \emph{splot} task in IRAF, instead, because the
Gaussian shape adopted by \emph{fitline} could not satisfactorily
reproduce the observed profile (e.g. very broad or asymmetric
lines). The equivalent widths used to determine the iron content range
approximately from 5 to 150~m\AA, well sampling the linear part of the
curve of growth. Full details on the selection of the iron lines and
their physical properties (oscillator strengths, etc.)  will be given
in Mottini et al (in preparation).

Following Kovtyukh \& Gorlova~(\cite{kov00}), we have determined the
stellar effective temperature from 32 line-depths ratios. Gravity and
microturbulent velocity were constrained by imposing the ionization
balance and by minimizing the slope of $\log(\epsilon(\mathrm{Fe}))$
vs EW, respectively. The LTE stellar model atmospheres by
Kurucz~(\cite{kur93}) and the WIDTH9 code (Kurucz \cite{kur93}) were
used throughout the analysis.  The detailed description of this
procedure is beyond the scope of this Letter and will also be
presented in Mottini et al (in preparation).

The mean value [Fe/H] is about solar for our Galactic sample,
$\sim-0.4$ for the LMC sample and $\sim-0.7$ for the SMC one, with an
rms of about 0.15~dex.

\subsection{Periods, photometry and distances\label{sec:phot}}
We have adopted the periods, distance moduli and $V$-band photometry
of the Galactic Cepheids as listed in Table~3 of Storm et
al~(\cite{sto04}).  Two of our programme stars, $\zeta~\mathrm{Gem}$
and $\beta~\mathrm{Dor}$, are not included in that list and for them
we have used the values from Groenewegen et al~(\cite{gro04b}, Table
3).

The periods and $V$-band photometry for the Magellanic Cloud Cepheids
were taken from Laney \& Stobie~(\cite{lan94}). The distance modulus
of the barycenter of the LMC is assumed to be 18.50, for consistency
with the $PL$ relation of Freedman et al~(\cite{fre01}, see
below). The SMC is considered 0.44 magnitudes more distant (e.g. Cioni
et al \cite{cio00}). Depth and projection effects in the Magellanic
Clouds were corrected for using the position angle and inclination of
each galaxy as determined by van der Marel \& Cioni~(\cite{mar01},
LMC) and Caldwell \& Laney~(\cite{cal91}, SMC).

The periods of our programme stars range between about 5 and 65 days,
thus populating the linear part of the $PL$ relation, the one useful
for distance determinations (e.g. Bono et al~\cite{bon99}). All of
stars are \emph{bona fide} fundamental mode pulsators. In particular,
the two stars with periods shorter than 8 days (\object{V Cen} and
\object{HV 6093}) follow the fundamental mode $PL$ relation of
Freedman et al~(\cite{fre01}), with deviations from it which are
consistent with its intrinsic scatter ($0.2$ and less than $-0.05$
magnitudes, respectively).

\section{The $PL$ relation vs [Fe/H]\label{sec:plz}}
\begin{figure*}[!ht]
  \centering
    \includegraphics[width=0.42\linewidth]{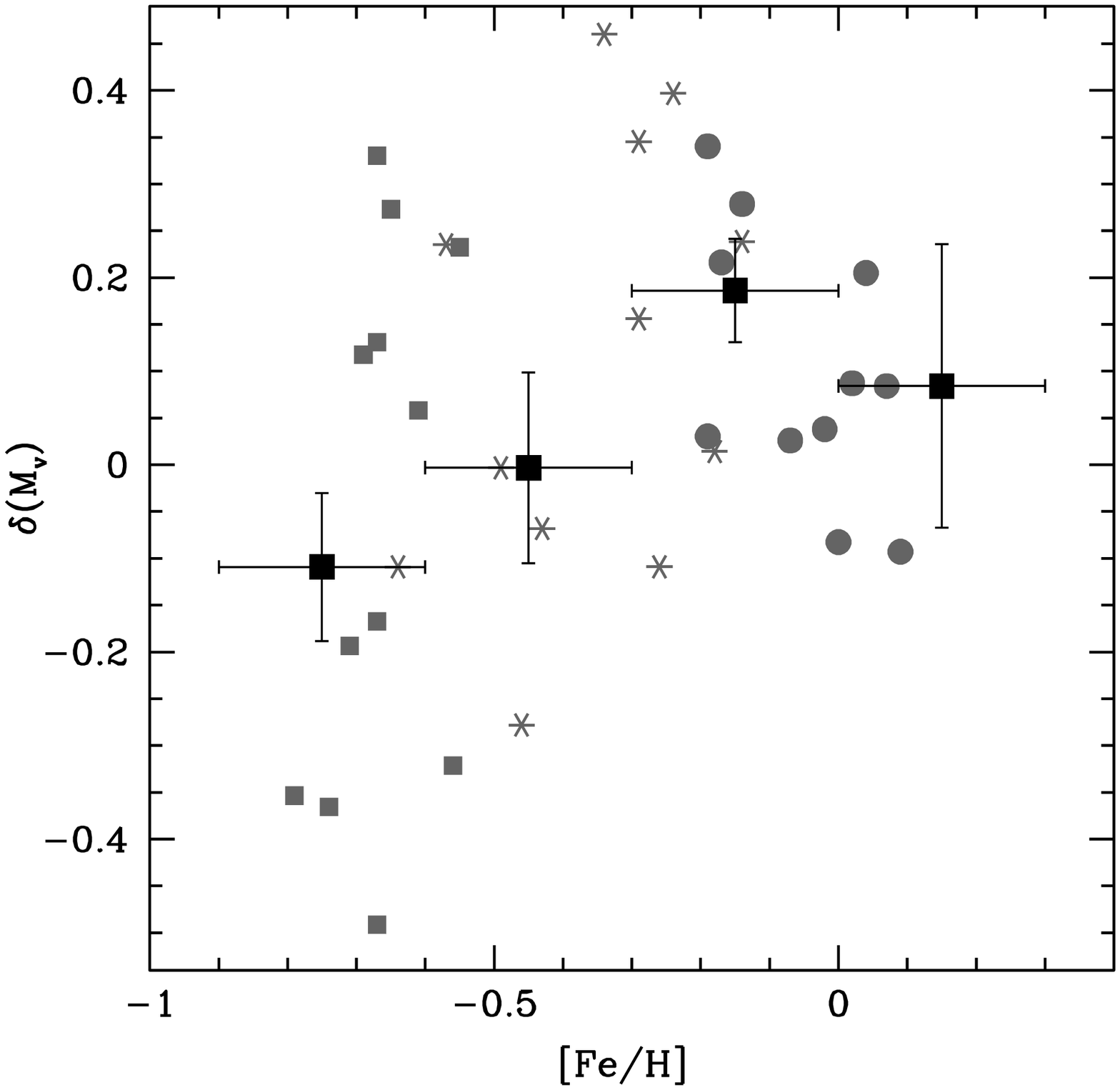}\hspace{\fill}
    \includegraphics[width=0.42\linewidth]{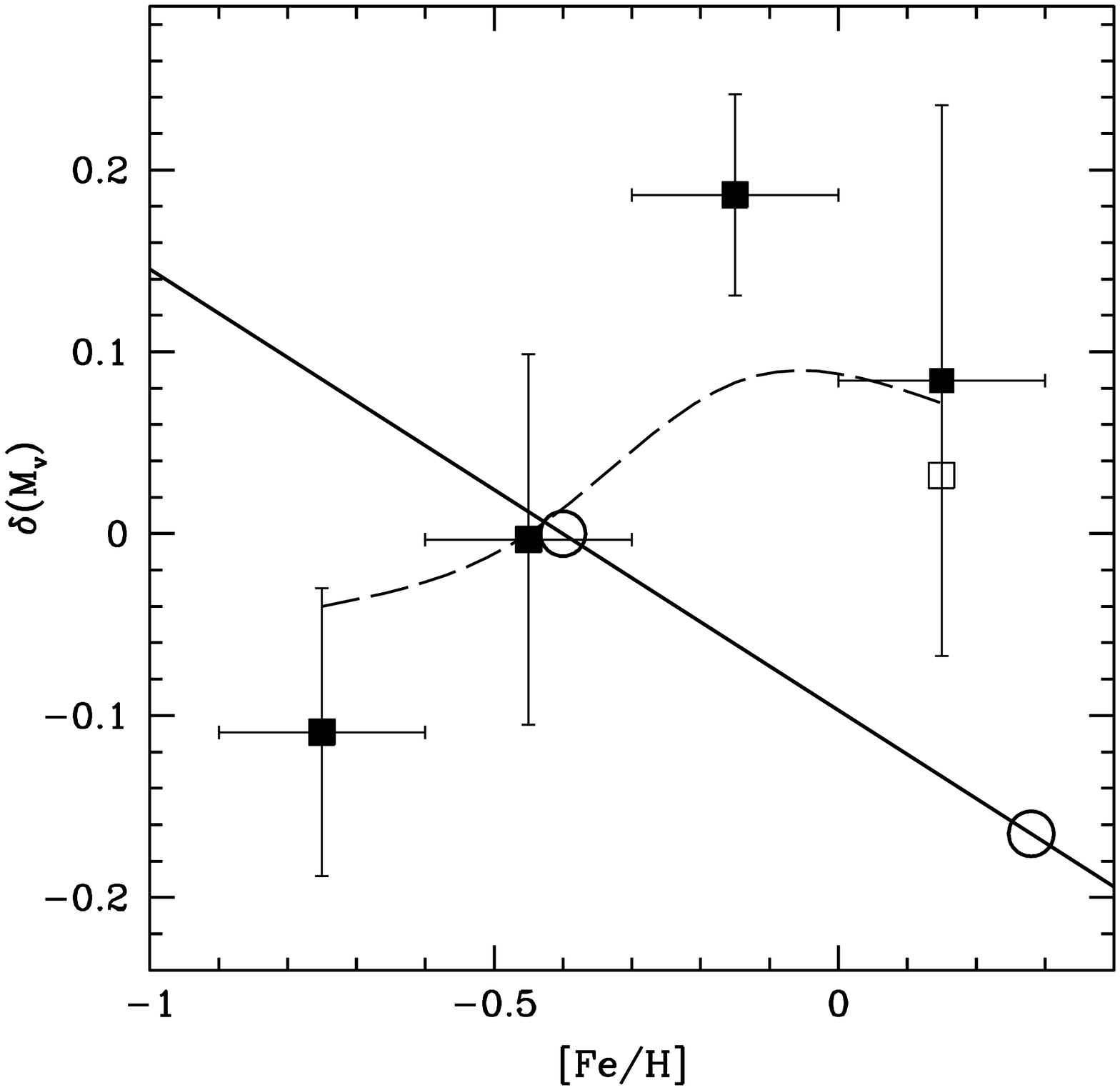}
    \caption{The $V$-band residuals compared to the Freedman et
    al~(\cite{fre01}) $PL$ relation are plotted against the iron
    content measured from FEROS and UVES spectra. \emph{Left panel:}
    results for the individual stars (grey symbols: Galaxy as circles,
    LMC as star symbols, SMC as squares) and median value in each
    metallicity bin (filled black squares) with its associated
    errorbar. \emph{Right panel:} Filled squares represent the median
    value in each metallicity bin, with is associated errorbar. The
    open square includes the three stars with metallicity from
    Andrievsky et al~(\cite{and02a,and02b}) and radius from Laney \&
    Stobie~(\cite{lan95}). The metallicity dependence as inferred by
    Kennicutt et al~(\cite{ken98}) from two Cepheid fields in M101
    (open circles) is shown as a full line. The dashed line shows the
    theoretical predictions by Fiorentino et al~(\cite{fio02}) for a
    helium-to-metal enrichment of $\Delta Y/\Delta Z=2.5$ and the
    observed period and $V-K$ distributions of the programme stars.}
    \label{fig:resz}
\end{figure*}

In Figure~\ref{fig:resz} we plot $\delta(M_{V})$, the $V$-band
residuals of our programme stars from the standard $PL$ relation of
Freedman et al~(\cite{fre01}), as a function of the iron abundance we
have derived from the FEROS and UVES spectra (a positive
$\delta(M_{V})$ means fainter than the mean relation). The Freedman et
al~(\cite{fre01}) $PL$ relation was derived for the LMC as a whole
$\left(\overline{\mathrm{[Fe/H]}}\simeq-0.4\right)$ and
$\delta(M_{V})$ is the correction to be applied to it as a function of
metallicity.

The data in Figure~\ref{fig:resz} are binned in metallicity to reflect
the typical uncertainty on our determination of [Fe/H], marked by the
horizontal errorbars. The median value of $\delta(M_{V})$ in each
metallicity bin is plotted as filled squares, with the vertical
errorbars representing its associated errors. The rms on
$\delta(M_{V})$ in each bin is of the order of 0.3~mag, corresponding
to the intrinsic width of the instability strip.

As it can be seen in Figure~\ref{fig:resz}, our data indicate that
$\delta(M_{V})$ increases with [Fe/H] up to about solar metallicity,
\emph{i.e.} Cepheids become fainter as metallicity increases, where it
shows a turnover or a flattening. Regrettably, the rather large
errorbars on $\delta(M_{V})$ do not allow us to distinguish between
these two possibilities. On the other hand, the data are only
marginally consistent with $\delta(M_{V})$ still rising at
metallicities higher than solar: a linear extrapolation of the trend
defined by the three points at lower metallicity would be $1.6\sigma$
away from the measured value at $\mathrm{[Fe/H]}=0.15$.

In order to enhance the statistics in the highest metallicity bin, we
have included in our sample 3 stars, \object{SZ Aql}, \object{WZ Sgr}
and \object{KQ Sco}, with published metallicities (Andrievsky et al
\cite{and02a,and02b}; $\mathrm{[Fe/H]}\simeq0.15$) and distances,
which we have derived by combining the radii measured by Laney \&
Stobie~(\cite{lan95}) with $K$-band photometry (Laney \& Stobie
\cite{lan94}) and the surface brightness calibration of Groenewegen
(\cite{gro04a}). The result is shown as an open square in
Figure~\ref{fig:resz}. With this addition, the fact that
$\delta(M_{V})$ keeps rising up to super-solar metallicity can be
excluded at the $3\sigma$ level.

\subsection{Comparison with previous results\label{sec:prev}}
Let us now use a $\chi^2$ technique to compare our data to selected
previous results:

\begin{itemize}
  \item {\bf null hypothesis}, \emph{i.e.} no dependence of
$\delta(M_{V})$ on [Fe/H]: the $\chi^2$ test returns a value of
13.5. For 3 degrees of freedom, this means that the hypothesis can be
discarded at the 99.6\% level. The data can be reconciled with the
null hypothesis if a distance modulus of 18.2-18.3 is used for the
LMC ($\chi^2\simeq3.1$ for 3 degrees of freedom). Such a short
distance, however, seems nowadays to be rather implausible (e.g.,
Romaniello et al \cite{rom00}, Walker \cite{wal03}, Alcock et al
\cite{alc04}, Borissova et al \cite{bor04})
  \item {\bf monotonically decreasing {\boldmath ${\delta(M_{V})}$}}
(e.g. Kennicutt et al \cite{ken98}, Sakai et al \cite{sak04}, Storm et
al \cite{sto04}, Groenewegen et al \cite{gro04b}): notably, this sort
of behavior was used in the final paper of the {\sc hst} Key Project
on $\mathrm{H}_0$ (Freedman et al~\cite{fre01}). As an illustration,
we report in Figure~\ref{fig:resz} the classical results of Kennicutt
et al~(\cite{ken98},
$\delta(M_{V})=(-0.24\pm0.16)\times\mathrm{[Fe/H]}$, open circles and
solid line). The $\chi^2$ test on it returns a value of 28, meaning
that it is incompatible with our data with a confidence higher than
99.95\%.
  \item {\bf theoretical models by Fiorentino et al~(\cite{fio02})}:
these are the outcome of non-linear pulsation computations and predict
a non-monotonic behavior of ${\delta(M_{V})}$ with [Fe/H].  Following
the prescriptions in that paper, we have computed the value of
${\delta(M_{V})}$ in each metallicity bin by combining the theoretical
Period-Luminosity-Color ($PLC$) relations as a function of metal
content (see their Table~4) with the observed periods and $V-K$ colors
of our stars. The use of the $PLC$ relations supplies individual
absolute magnitudes, thus avoiding deceptive uncertainties due to the
intrinsic width of the instability strip. In the left panel of
Figure~\ref{fig:resz} we report as a dashed line the results for the
models with a helium-to-iron enrichment of $\Delta Y/\Delta
Z=2.5$. The $\chi^2$ value for this model is 4.2 which, for 3 degrees
of freedom, indicates a fairly good agreement with the data.  The
other two models presented in Fiorentino et al~(\cite{fio02}), $\Delta
Y/\Delta Z=3\ \mathrm{and}\ 3.5$, also result in similarly acceptable
values of $\chi^2$.
\end{itemize}

\section{Discussion and conclusions\label{sec:disc}}
We have assessed the influence of the stellar iron content on the
Cepheid Period-Luminosity ($PL$) relation by relating the $V$ band
residuals from the Freedman et al~(\cite{fre01}) $PL$ relation to
[Fe/H] for 37 Galactic and Magellanic Clouds Cepheids. The iron
abundance was measured from FEROS and UVES high-resolution and
high-signal to noise optical spectra. The novelty of our approach
consists in measuring directly the chemical composition of Cepheid
stars with a known distance, without relying on proxies such as oxygen
nebular abundances derived from spectra of H~II regions at the same
galactocentric distance of the Cepheid field (e.g. Kennicutt et al
\cite{ken98}), or secondary distance indicators like the Tip of the
Red Giant Branch (e.g. Sakai et al \cite{sak04}). Here we have
concentrated on the influence of the iron content, but the very same
analysis can be applied to any other chemical species (Mottini et al,
in preparation).

Our results are summarized in Figure~\ref{fig:resz} (filled squares),
together with the empirical results of Kennicutt et al~(\cite{ken98})
in two Cepheid fields in M101 (open circles and solid line) and the
theoretical predictions by Fiorentino et al~(\cite{fio02}) from
non-linear pulsational models (dashed line).  The data indicate that
$\delta(M_{V})$, the correction to a metal-independent $PL$ relation,
increases as the iron content increases, \emph{i.e.}  the stars become
fainter as the metallicity increases, until a flattening or a turnover
is reached at about solar metallicity. The possibility that the
increasing trend continues at higher metallicities, while it cannot be
ruled out completely by the current data, is disfavored at the
$1.6\sigma$ level.

Applying a $\chi^2$ technique, the null hypothesis, \emph{i.e.}  no
dependence of the $PL$ relation on the iron content, can be excluded
at the 99.6\% level. Also, empirical monotonically decreasing linear
relations (e.g. Kennicutt et al \cite{ken98}, Sakai et al
\cite{sak04}, Storm et al \cite{sto04}, Groenewegen et al
\cite{gro04b}) are incompatible with our data with a confidence level
higher than 99.95\%. A better agreement is found with the theoretical
models of Fiorentino et al~(\cite{fio02}), which do predict a
non-monotonic behavior with a turnover at about solar metallicity. The
$\chi^2$ value is 4.2 (3 degrees of freedom) for their model with
$\Delta Y/\Delta Z=2.5$, plotted as a dashed line in
Figure~\ref{fig:resz}. A similarly good agreement is found for
their models with $\Delta Y/\Delta Z=3$ and $3.5$.

It is apparent from an inspection of Figure~\ref{fig:resz} that,
because of the limited number of stars, the errorbars on the data are
quite large. However, the error on $\delta(M_{V})$ scales as the
square root of the number of stars in each bin. It is, then, just a
matter of gathering a larger sample of stars and apply the method we
have outlined here to characterize in detail the dependence of the
Cepheid $PL$ relation on the stellar chemical composition.

\begin{acknowledgements}
We warmly thank Emanuela Pompei for carrying out part of the FEROS
observations for us. We would also like to thank the ESO staff for
successfully executing our UVES observations in Service Mode.  Several
stimulating discussions with Ferdinando ``Nando'' Patat are gratefully
acknowledged. G.B. acknowledges financial support by INAF2002 under
the project ``The Large Magellanic Cloud as a laboratory for stellar
astrophysics''.

\end{acknowledgements}

\begin{table}
\caption{Selected properties of the programme Cepheid stars.  Iron
abundances were measured from our FEROS and UVES spectra (see
section~\ref{sec:detz}). Periods, distance moduli and $V$-band
photometry of the Galactic Cepheids are taken from Storm et
al~(\protect{\cite{sto04}}), except $\zeta~\mathrm{Gem}$ and
$\beta~\mathrm{Dor}$ from Groenewegen et al~(\protect{\cite{gro04b}}).
The periods and $V$-band photometry for the Magellanic Cloud Cepheids
were taken from Laney \& Stobie~(\protect{\cite{lan94}}). The distance
modulus of the barycenter of the LMC is assumed to be 18.50 (see
text). The SMC is considered 0.44 magnitudes more distant (e.g. Cioni
et al \protect{\cite{cio00}}). Depth and projection effects in the
Magellanic Clouds were corrected for using the position angle and
inclination of each galaxy as determined by van der Marel \&
Cioni~(\protect{\cite{mar01}}, LMC) and Caldwell \&
Laney~(\protect{\cite{cal91}}, SMC).  }

\centering
\begin{tabular}{llcc}
\hline\hline
   Name & log(P) & $M_{V}$  & [Fe/H]\\
\hline
  \multicolumn{4}{c}{Galaxy}    \\
  V  Cen & 0.740 & -3.295 & +0.04 \\
 KN  Cen & 1.532 & -6.328 & +0.17 \\  
 VW  Cen & 1.177 & -4.037 & -0.02 \\	  
  S  Nor & 0.989 & -4.101 & +0.02 \\	  
  T  Mon & 1.432 & -5.372 & -0.02 \\	  
  U  Nor & 1.102 & -4.415 & +0.07 \\	
 AQ  Pup & 1.479 & -5.513 & -0.07 \\
 VZ  Pup & 1.365 & -5.009 & -0.17 \\
 RS  Pup & 1.617 & -6.015 & +0.09 \\
 RZ  Vel & 1.310 & -5.042 & -0.19 \\
  l  Car & 1.551 & -5.821 &  0.00 \\
beta Dor & 0.993 & -3.920 & -0.14 \\   
zeta Gem & 1.007 & -3.897 & -0.19 \\   
\hline
  \multicolumn{4}{c}{LMC}   \\
HV  1013 & 1.382 & -5.037 & -0.57 \\	 
HV  1023 & 1.425 & -4.994 & -0.24 \\	 	 
HV 12452 & 0.941 & -3.899 & -0.29 \\	 	 
HV 12700 & 0.911 & -3.627 & -0.29 \\	 	 
HV  2260 & 1.112 & -4.067 & -0.34 \\	 	 	
HV  2294 & 1.563 & -6.050 & -0.46 \\	 	 
HV  2352 & 1.134 & -4.656 & -0.43 \\	 	 
HV  2369 & 1.684 & -6.215 & -0.64 \\	   
HV   997 & 1.119 & -4.308 & -0.14 \\	 	 		
HV  6093 & 0.680 & -3.338 & -0.49 \\	 	 
HV  2580 & 1.228 & -4.833 & -0.18 \\	 	 
HV  2733 & 0.941 & -4.164 & -0.26 \\	 	 
\hline
  \multicolumn{4}{c}{SMC}   \\
HV 11211 & 1.330 & -5.296 & -0.67 \\
HV 1365  & 1.094 & -4.147 & -0.67 \\
HV 1954  & 1.223 & -5.325 & -0.67 \\
HV 2064  & 1.527 & -5.440 & -0.55 \\
HV 2195  & 1.621 & -5.874 & -0.61 \\
HV 2209  & 1.355 & -5.519 & -0.56 \\
HV 817   & 1.277 & -5.348 & -0.74 \\
HV 823   & 1.504 & -5.336 & -0.65 \\
HV 824   & 1.818 & -6.669 & -0.71 \\
HV 837   & 1.631 & -5.842 & -0.69 \\
HV 847   & 1.433 & -5.282 & -0.67 \\
HV 865   & 1.523 & -6.015 & -0.79 \\
\hline
\end{tabular}
\end{table}

\end{document}